\documentclass[10pt,letterpaper]{article}
\usepackage{opex3,bm,cite}

\begin{document}

\title{Near-perfect absorption in epsilon-near-zero structures with hyperbolic dispersion}

\author{Klaus Halterman$^{1,*}$ and J. Merle Elson$^2$}
\address{$^1$Michelson Lab,
Physics Division, Naval Air Warfare Center,  China Lake, California
93555, USA \\ 
$^2$PO Box 965,  Tijeras, New Mexico 87059, USA}
\email{$^*$klaus.halterman@navy.mil} 


\begin{abstract}
We investigate the interaction of polarized
electromagnetic waves
with hyperbolic metamaterial
structures, 
whereby the 
in-plane permittivity component $\epsilon_x$ 
is opposite in sign to the normal
component $\epsilon_z$.
We find 
that when the thickness of the metamaterial is 
smaller than the wavelength of  the incident wave, 
hyperbolic metamaterials can absorb significantly higher amounts of
electromagnetic energy compared to their conventional counterparts.
We also demonstrate
that for wavelengths leading to $\Re(\epsilon_z) \approx 0$,
near-perfect absorption arises and
persists over 
a range of frequencies and subwavelength structure thicknesses.
\end{abstract}



\section{Introduction}
With recent advances in nanoscale fabrication of
metal-dielectric multilayers and arrays of rods, 
hybrid structures can now be created
that
absorb a 
substantial portion of incident electromagnetic (EM) radiation. 
In conventional approaches, strong absorption
was achieved by utilizing materials that had either
high loss or large thickness.
Nowadays, with the advent of metamaterials,
absorbing structures can be created
that
harness  
plasmonic excitations or implement
high impedance
components \cite{siev} that have  extreme values  \cite{enghetta,klaus,klaus2,klaus3}  of the 
permittivity $\bm \epsilon$  or  permeability $\bm \mu$. 
In close
connection with these developments,
there has also been a substantial amount of research lately involving anisotropic 
metamaterials, where now
$\bm \epsilon$
and $\bm \mu$ are tensors that have
in general differing components along the three coordinate axes.
An important type of anisotropic metamaterial
is one whose corresponding orthogonal 
tensor components are of 
opposite sign, sometimes referred to as indefinite media \cite{shurig}.
When such structures are described by a diagonal tensor,
the corresponding dispersion relation permits wavevectors that lie within a hyperbolic isofrequency surface, and hence 
such a material is also called a hyperbolic metamaterial (HMM).
The inclusion of HMM elements in many designs can be beneficial due to their inherent nonresonant character, thus 
limiting loss effects \cite{cortes}. 

The earliest HMM construct involving bilayers of anisotropic media was discussed 
in the context of bandpass spatial filters with tunable cutoffs \cite{shurig}.
For wavelengths $\lambda$ in the visible spectrum, an effective HMM was modeled
using arrays of metallic nanowires \cite{liu}  spaced apart distances
much smaller than $\lambda$, thus
avoiding the usual problems associated with resonances.
Periodic arrays of carbon nanotubes \cite{nef} have been shown to
exhibit HMM characteristics 
in the THz spectral range.
Other possibilities involve metal-dielectric layers:
The inclusion of active media  
in  metal-dielectric multilayers
can result in improved HMM-based imaging devices \cite{ni}.
For certain layer configurations,
nonlocal effects \cite{sav}, which depending on geometry \cite{kid}, can
limit the number of accessible photonic states \cite{yan2}.
The absorption in thin films has been shown experimentally to be enhanced when in contact with a multilayered HMM substrate \cite{tum}.
Rather than 
using metallic components, tunable graphene can 
switch between a hyperbolic and conventional material via a gate voltage \cite{orsh}.
The HMM dispersion can be tuned in gyromagnetic/dielectric \cite{li} 
and 
semiconductor/dielectric structures \cite{rizza}.
Slabs of semiconductors can also exhibit tunability
by photogenerating a grating
via variations in  the carrier density caused by two incident beams,  
revealing a hyperbolic character \cite{rizza2}.

\begin{figure}
\centerline{\includegraphics[width=0.45\paperwidth]{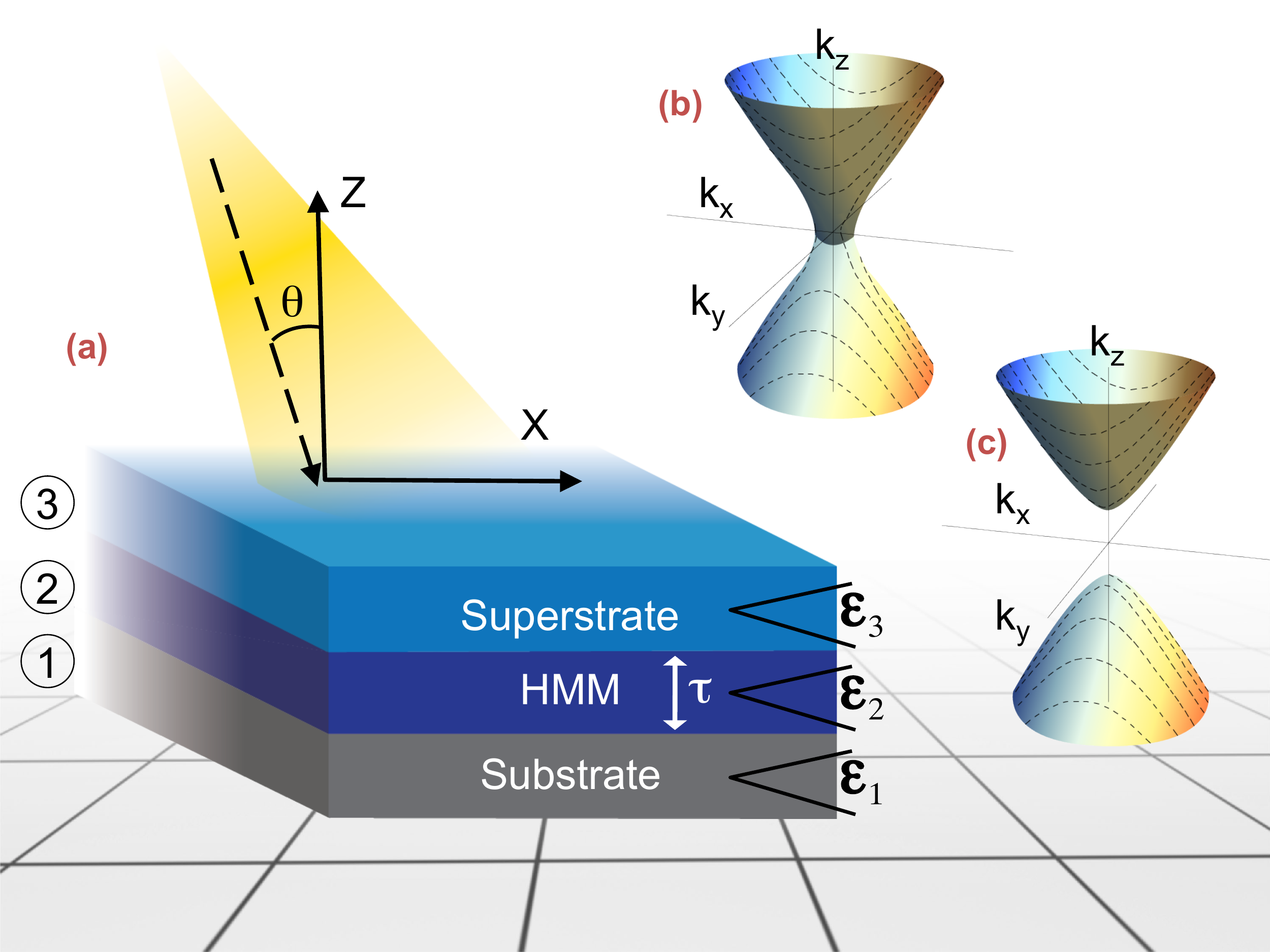}}
\caption{(a) Schematic of the hyperbolic metamaterial configuration: 
The  HMM layer of thickness 
$\tau$ is bordered by a semi-infinite superstrate and substrate. The
permittivites ${\bm \epsilon}_i$ $(i$$=$$1,2,3)$ are in general anisotropic.
The incident field is polarized in the $x$$-$$z$ plane at an
angle $\theta$. (b) 
Dispersion contours for a type-2 HMM where $\epsilon_x$$<$$0$, and $\epsilon_z$$>$$0$
and (c) for a type-1 HMM with $\epsilon_x$$>$$0$, and $\epsilon_z$$<$$0$.}   
\label{fig1}
\end{figure}
Increased absorption can also be achieved by 
incorporating a grating 
with the HMM, so that by introducing
surface corrugations, 
or grooves,
light can diffract and generate a broad spectrum of wave vectors 
into the HMM layer.
These wavevectors  can couple via surface modes \cite{yan} due to the impedance 
mismatch at the various openings.
Grating lines were patterned above a layered ${\rm Au /TiO_2}$ HMM structure, creating a
``hypergrating" capable of exciting both surface and bulk plasmons \cite{sreek}.
By judiciously designing the materials below the grating, 
it can be possible to absorb a considerable fraction of the diffracted EM field.
Indeed, a HMM comprised of arrays of silver nanowires was experimentally shown to
reduce the reflectance  by introducing surface corrugations \cite{nari}.
Spherical nanoparticles deposited on  planar HMM structures
also resulted in reduced reflectance due to the 
increased density of photonic states \cite{tumkur}.

In this paper we 
show that near-perfect absorption
of EM radiation can arise in a simple
HMM structure adjacent to a
metal.
We 
investigate
a range of frequencies where the 
permittivity components
perpendicular and parallel to the interfaces are of opposite sign.
We consider 
two possibilities: when the HMM dispersion relation is 
of type-1 or type-2, which for our geometry corresponds to
$\epsilon_x$ $>$ $0$,  $\epsilon_z$ $<$ $0$ or 
$\epsilon_x$ $<$ $0$,  $\epsilon_z$ $>$ $0$
respectively (see Fig.~\ref{fig1}).
We show that for those $\lambda$ leading to
the real part of the permittivity component perpendicular to the interfaces ($\epsilon_z$)
nearly vanishing,  an  intricate balance between material loss
and structure thickness ($\tau$) yields
a broad range of incident angles $\theta$ and
$\tau$ 
in which nearly the entire EM wave is absorbed.
These findings are absent in  conventional anisotropic  ``elliptical" structures. 

\section{Methods}
\begin{figure*}
\centerline{\includegraphics[width=.34\paperwidth]{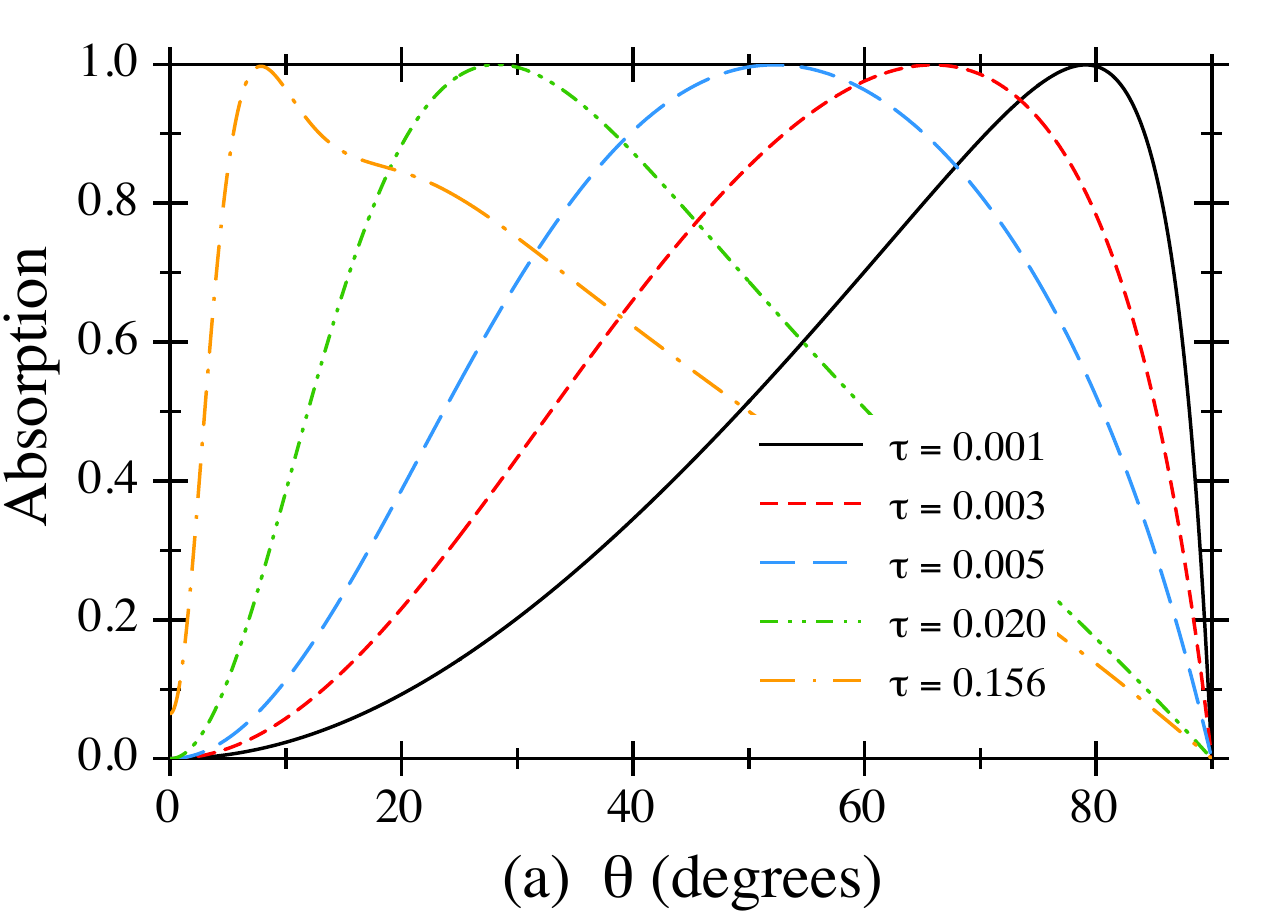} 
\includegraphics[width=.34\paperwidth]{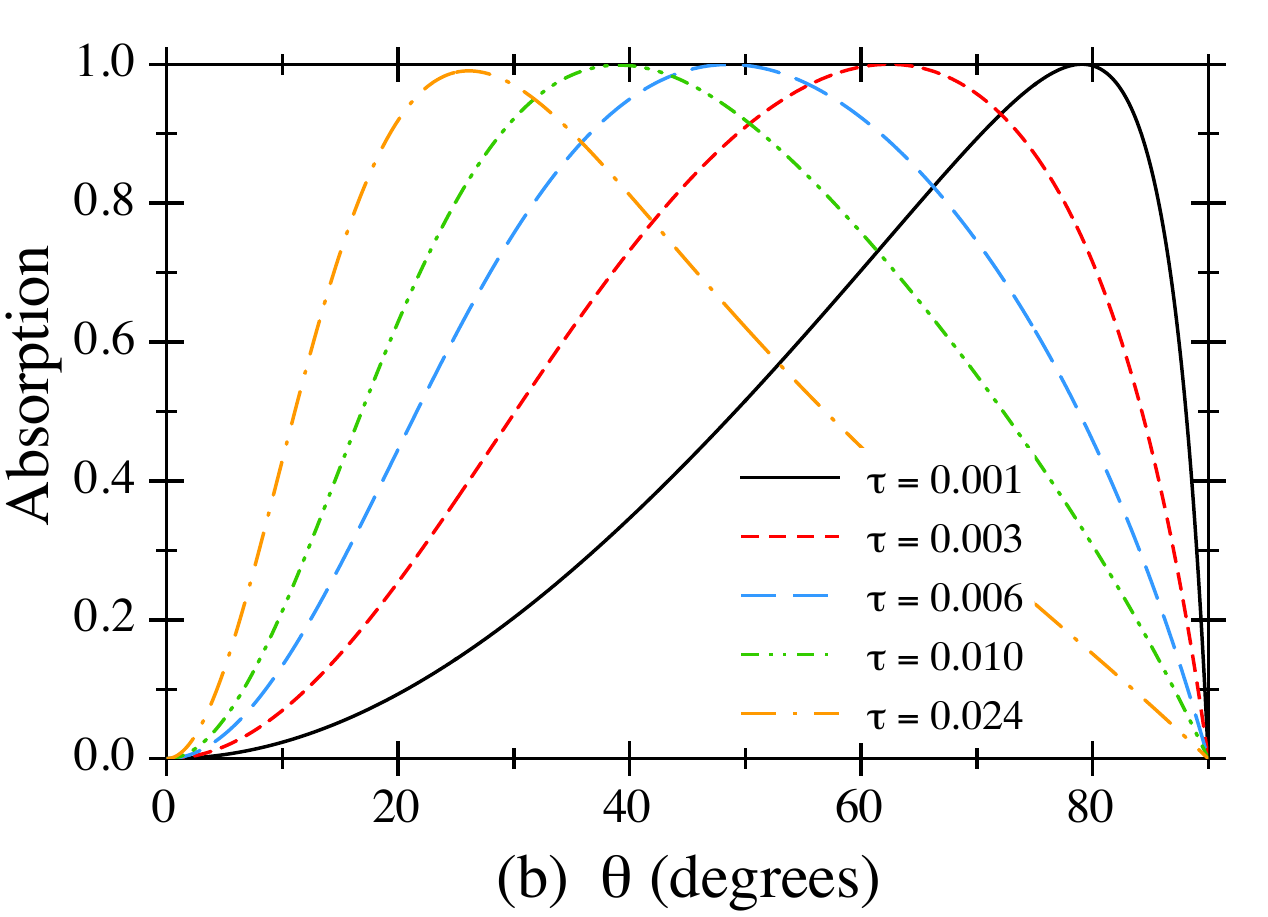} } 
\vspace{-.1 cm}
\centerline{\includegraphics[width=.34\paperwidth]{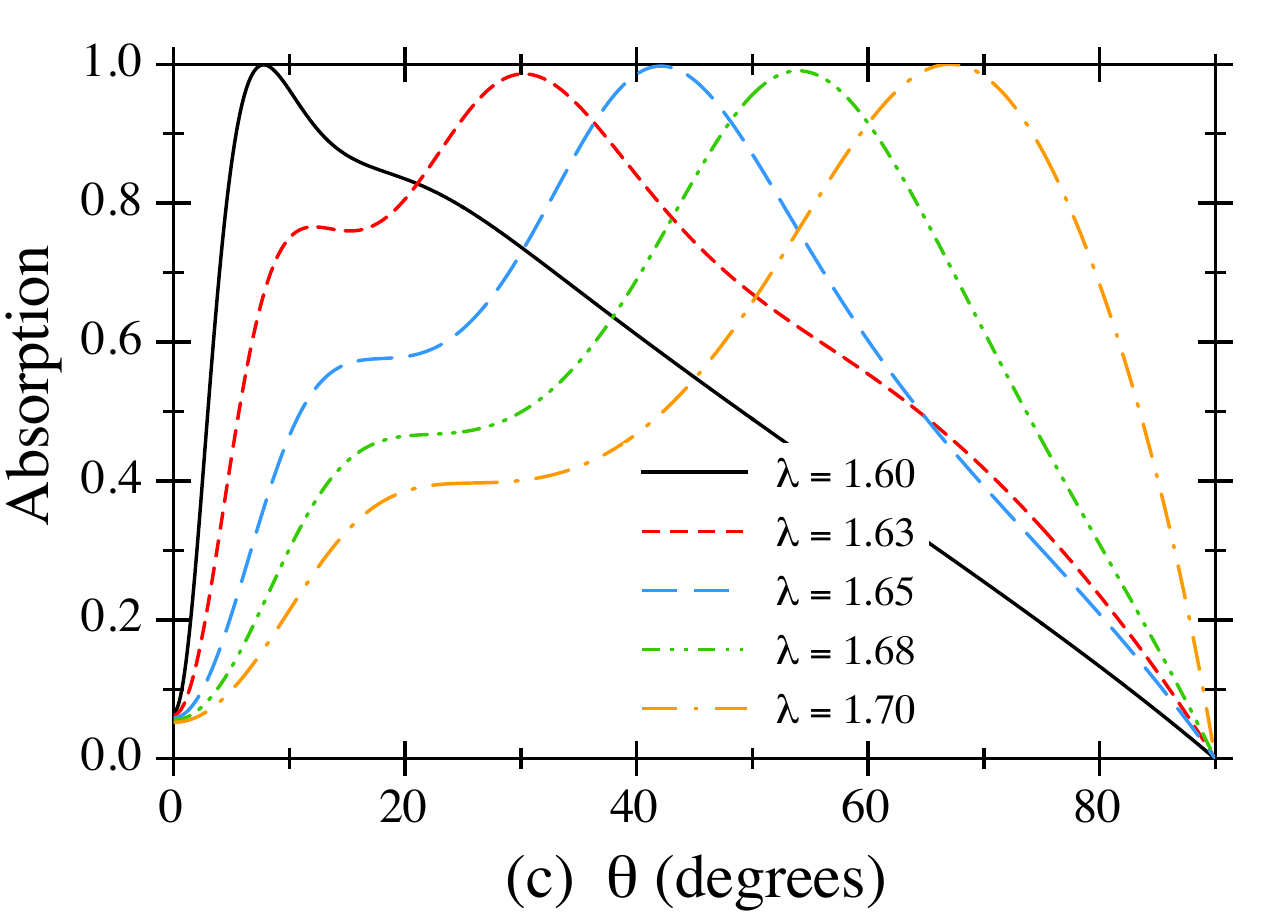} 
\includegraphics[width=.34\paperwidth]{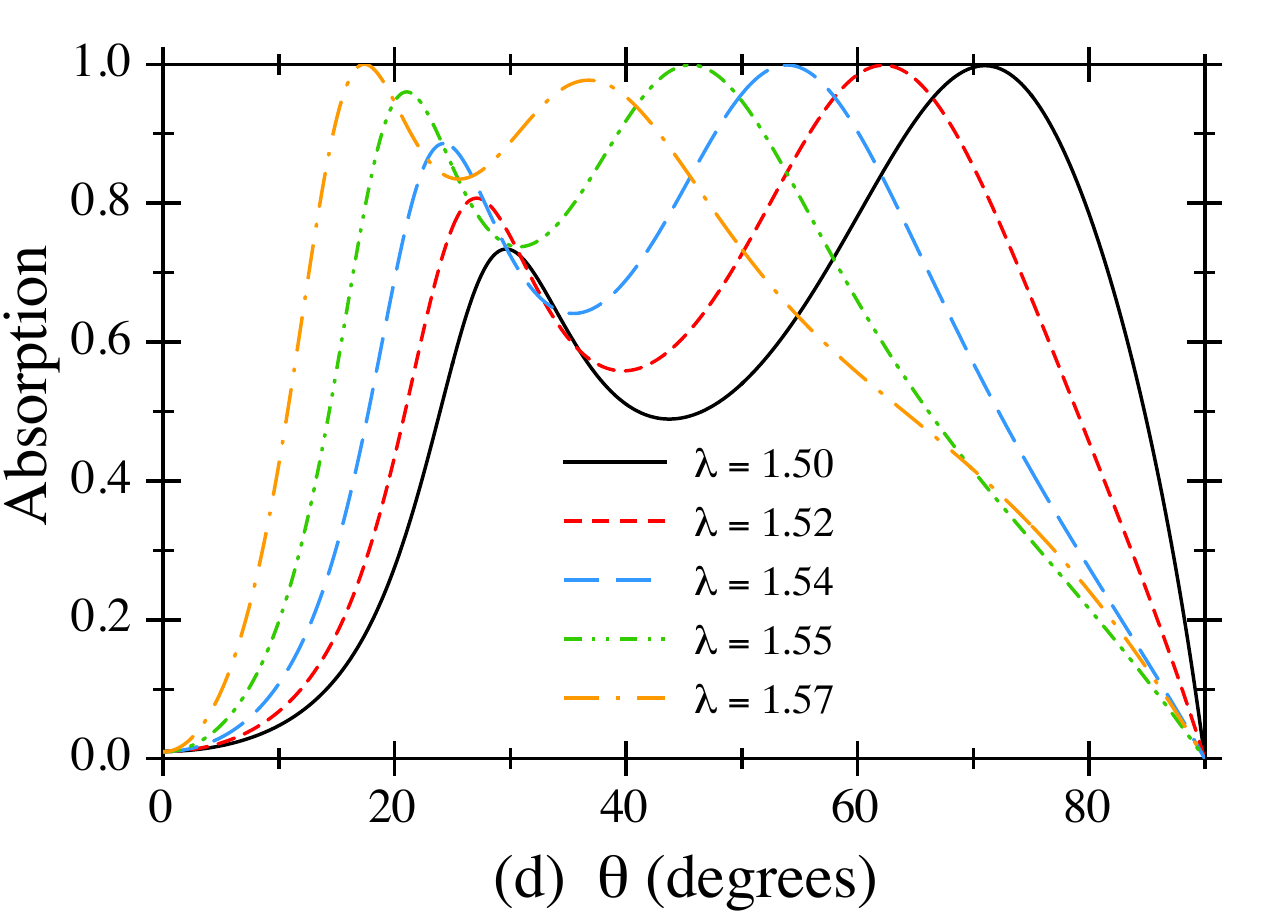} }
\caption{Absorption 
as a function of incident angle $\theta$. 
The superstrate is air, and the HMM layer is supported by a perfectly conducting substrate.
In 
(a) and (b) a
range of HMM widths $\tau$ are studied (legend units are in microns). 
In (a) $\Re(\epsilon_{2x})>0$ and $\Re(\epsilon_{2z})< 0$ (type-1 HMM),
and in 
(b)  $\Re(\epsilon_{2x})<0$, 
and $\Re(\epsilon_{2z})>0$ (type-2 HMM). For
both panels (a) and (b), 
$\lambda \approx \lambda_z$
so that $\Re(\epsilon_{2z}) \approx 0$.
Panels (c) and (d) show the effects of varying $\lambda$
for both  the type-1 and type-2 cases respectively.  For those 
cases
$\tau$ is fixed at $0.16$ $\mu$m.
For normal incidence ($\theta=0^\circ$), there is generally
little absorption (high reflectance). 
Remarkably, for a range of HMM widths 
and wavelengths
there are strong absorption peaks spanning a broad range of $\theta$. 
}  
\label{fig2}
\end{figure*}

We assume that the 
incident EM wave propagates with wave vector in the $x-z$ plane with polarization 
($E_x, E_z, B_y$) ($p$-polarized) or ($E_y, B_x, B_z$) ($s$-polarized). 
Once the wave enters the anisotropic medium, its polarization
state can then split into linear combinations of both TE and TM polarizations \cite{zhao}.
Consider an unbounded diagonally anisotropic medium described by homogeneous parameters 
($\epsilon_x, \epsilon_y, \epsilon_z$) and ($\mu_x, \mu_y, \mu_z$),
where it is always possible to choose principal coordinate axes so that the permittivity and permeability are diagonal.
Assuming a harmonic time dependence, 
$\exp(-i \omega t)$,
for the EM fields, 
Maxwell's equations give
the corresponding wave equations for the electric field components $E_x$ and $E_y$:
\begin{eqnarray}
\label{e11a}
 \frac{\partial^2 E_x }{\partial z^2} + \left[\left(\frac{\omega}{c}\right)^2 \epsilon_{x}\mu_{y} - \left(\frac{\epsilon_{x}}{\epsilon_{z}}\right) k_x^2 \right]  E_x = 0, \\
\label{e11b}
 \frac{\partial^2 E_y}{\partial z^2} + \left[\left(\frac{\omega}{c}\right)^2\epsilon_{y}\mu_{x}-\left(\frac{\mu_{x}}{\mu_{z}}\right)k_x^2 \right] E_y = 0.
\end{eqnarray}
Equations (\ref{e11a}) and (\ref{e11b}) illustrate that the wave equations are 
different for $E_x$ and $E_y$, resulting in two different wave vectors.
In this work, we focus exclusively on $p$-polarization  from which the nature of the 
HMM dispersion can be qualitatively understood.
From  Eq.~(\ref{e11a}),
${\hat k}_z^2 = \epsilon_{x}\mu_{y} - \left({\epsilon_{x}}/{\epsilon_{z}}\right) {\hat k}_x^2$
(the caret symbol signifies normalization by $\omega/c$). For this discussion
we assume real valued material parameters and positive $\mu_{y}$. Focusing on $\epsilon_{x}$$>$$0$,
we consider two scenarios
(a) $\epsilon_{z}$$ >$$ 0$ and (b) $\epsilon_{z}$$ <$$ 0$, yielding the respective dispersion relations
${{\hat k}_z^2}/({ \epsilon_{x}\mu_{y}}) + {{\hat k}_x^2}/({\epsilon_{z}\mu_{y}})  = 1$
and ${{\hat k}_z^2}/({ \epsilon_{x}\mu_{y}} )
- {{\hat k}_x^2}/({|\epsilon_{z}|\mu_{y}}) = 1$.
Thus the isofrequency contours are (a) ellipses and (b) 
hyperbola (see e.g., Fig.~\ref{fig1}(c) when $k_y=0$).
Moreover, for the ellipsoidal case, as ${\hat k}_x$ increases there will be a frequency cutoff since ${\hat k}_z^2$ eventually becomes negative. 
On the other hand, for the hyperbolic case, when ${\hat k}_x$ increases, there is no cutoff since ${\hat k}_z^2$ remains positive.  
If $\epsilon_{x}<0$ and $\epsilon_z>0$, we then have the possibility of a connected hyperbola (see Fig.~\ref{fig2}(b)).

To determine the absorbed EM energy,
it is convenient to first determine the Fresnel reflection coefficient, $r$.
The corresponding reflectance $R$ 
is  then given by $R = |r|^2$.
For a $p$-polarized plane wave 
incident at an angle $\theta$ relative to the normal of a planar layer of thickness $\tau$, we find,
\begin{eqnarray}
\label{e15}
r = \beta \left[\frac
{({\hat k}_{z1}\epsilon_{x2}-{\hat k}_{z2}\epsilon_{x1})({\hat k}_{z2}\epsilon_{x3}+{\hat k}_{z3}\epsilon_{x2})e^{i\phi_2}+({\hat k}_{z1}\epsilon_{x2}+{\hat k}_{z2}\epsilon_{x1})({\hat k}_{z2}\epsilon_{x3}-{\hat k}_{z3}\epsilon_{x2})e^{-i\phi_2}}
{({\hat k}_{z1}\epsilon_{x2}-{\hat k}_{z2}\epsilon_{x1})({\hat k}_{z2}\epsilon_{x3}-{\hat k}_{z3}\epsilon_{x2})e^{i\phi_2}+({\hat k}_{z1}\epsilon_{x2}+{\hat k}_{z2}\epsilon_{x1})({\hat k}_{z2}\epsilon_{x3}+{\hat k}_{z3}\epsilon_{x2})e^{-i\phi_2}}\right],
\end{eqnarray}
where the semi-infinite substrate and superstrate are in general anisotropic (see Fig.~\ref{fig1}).
The details can be found in Sec.~\ref{app}.
We define, 
\begin{equation}
\label{beta}
\beta = \exp({-2i\phi_3}), 
\end{equation}
where
\begin{equation}\label{phij}
\phi_j \equiv (\omega/c){\hat k}_{zj}\tau,
\end{equation}
and
\begin{equation} \label{kzj}
{\hat k}_{zj}^2 \equiv \epsilon_{xj}\mu_{yj} - \left(\epsilon_{xj}/\epsilon_{zj}\right){\hat k}_x^2. 
\end{equation}
The 
index $j$ labels the regions $1,2$ or $3$ (see Fig.~\ref{fig1}).  
In all  cases below, the 
incident beam is in vacuum (region 3) so
that ${\hat k}_x = \sin\theta$, 
which is  conserved across the interface.
The frequency dispersion in the HMM takes the Drude-like form:
$\epsilon_{z2} = a+ib$,
where
$a = 1 - \alpha^2/[1+(\alpha f)^2]$,  and
$b = \alpha^3 f/[1+(\alpha f)^2]$. 
Here, $\alpha \equiv \lambda/\lambda_z$, 
$f=0.02$, and the characteristic wavelength, $\lambda_z = 1.6$$\mu$m.
When discussing the two types of HMM, 
the 
permittivity 
parallel to the interface is described
using $\epsilon_{x2} = \pm 4+0.1 i$
for type-1 ($+$) and type-2 ($-$).
The wavelength range considered here, where the system 
exhibits  HMM behavior  is 
consistent with experimental work involving HMM semiconductor hybrids \cite{naik}.
 
When the surrounding media is air and the central layer is a diagonally anisotropic HMM,
setting the numerator of (\ref{e15}) to zero leads
to a set of conditions on the wavevector components that
results in a complete absence of reflection  ($R = 0$):
\begin{eqnarray}
\label{d03.1}
{\hat k}_x^2 = \ \epsilon_{z2}\left(\frac{\epsilon_{x2}-\mu_{y2}}{\epsilon_{z2}\epsilon_{x2}-1}\right);\quad {\hat k}_{z2}^2 =  \epsilon_{x2} ^2\left(\frac{\epsilon_{z2}\mu_{y2}-1}{\epsilon_{z2}\epsilon_{x2}-1}\right) \\
\label{d03.2}
{\hat k}_x^2 = \ \epsilon_{z2}\mu_{y2}-\left(\frac{\epsilon_{z2}}{\epsilon_{x2}}\right)\left(\frac{n \lambda}{2 \tau}\right)^2;\quad {\hat k}_{z2}^2 
= \left(\frac{n \lambda}{2 \tau}\right)^2,
\end{eqnarray}
where 
$n$ is an integer. The $\hat{k}_x$ in Eq.~(\ref{d03.1}) corresponds to the classic 
Brewster angle condition for isotropic media:
${\hat k}_z^2\epsilon_{x2}^2 = {\hat k}_{z2}^2$, 
and
the $\hat{k}_x$ in Eq.~(\ref{d03.2}) corresponds to a standing wave condition 
in the $z$-direction. 
In either case, when Eqs.~(\ref{d03.1}) or (\ref{d03.2}) is satisfied, 
a minimum 
in $R$ arises. 
Under the Brewster angle condition in Eq.~(\ref{d03.1}), a simple
rearrangement shows that ${\hat k}_x^2 (\epsilon_{x2}-\epsilon_{z2}^{-1} )$ $=$ $\epsilon_{x2}-\mu_{y2}$.
This implies that if we choose $\epsilon_{x2}=\mu_{y2}$ and $\epsilon_{x2}\epsilon_{z2}=1$, then we should have $R=0$ for any value of $\hat{k}_x = \sin\theta$. 
This choice of anisotropic material parameters 
is similar to the perfectly matched layer (PML) approach 
to eliminating unwanted reflection from absorbing computational domain boundaries, 
especially in time-domain \cite{r08,r10}
and frequency-domain algorithms \cite{r11,r13}. 
Note that such a PML medium is 
somewhat artificial  
since $\epsilon_{x2}=\epsilon_{z2}^{-1}$ implies sources in region 2.  Nonetheless
such a concept is successful for absorbing layers designed to simulate an infinite computational domain. 

We now illustrate the important case of a
HMM backed by a perfectly conducting metal,
and the near-perfect absorption that can arise.
As is appropriate for HMM structures,
we also consider the regime where all materials are nonmagnetic (${\bm \mu}= 1$).
The reflection coefficient in Eq.~(\ref{e15}) then becomes,
\begin{eqnarray}
\label{d04}
r = e^{-2i\phi} \left[ \frac{
 ({\hat k}_{z2}+{\hat k}_{z}\epsilon_{x2}) e^{i\phi_2}-({\hat k}_{z2}-{\hat k}_{z}\epsilon_{x2}) e^{-i\phi_2} } 
 {({\hat k}_{z2}-{\hat k}_{z}\epsilon_{x2}) e^{i\phi_2}-({\hat k}_{z2}+{\hat k}_{z}\epsilon_{x2})e^{-i\phi_2} } \right],
\end{eqnarray}
where 
$
\phi \equiv (\omega/c){\hat k}_z \tau,
$
and  
$
{\hat k}_z = \cos\theta.
$
It is readily verified that for lossless media, Eq.~(\ref{d04}) yields perfect reflection ($|r|^2$$=$$1$) as expected.
In the absence of transmission, the 
absorption, $A$, is  simply written as  $A=1-R$.

\section{Results}
Figure~\ref{fig2} shows the absorption as a function of incident angle $\theta$ for
both types of HMM: type-1, $\Re \lbrace \epsilon_{2x} \rbrace >0$, $\Re \lbrace \epsilon_{2z} \rbrace
<0$ (panels a and c), and
type-2, $\Re \lbrace \epsilon_{2x} \rbrace<0$, $\Re \lbrace \epsilon_{2z} \rbrace>0$ (panels b and d).
Since  $\lambda \approx \lambda_z$, we have also the condition, $\Re(\epsilon_{2z}) \approx 0$. 
There cannot be any substrate transmission and thus 
$R<1$ is due to  intrinsic HMM losses. 
In terms of  practical designs, it is important to
determine the range of sub-wavelength HMM layer thicknesses that can admit perfect absorption. Thus
Figs.~\ref{fig2}(a) and (b) explore differing $\tau$  ranging from 0.001 to 0.156 $\mu$m.
Although the relative sign of $\epsilon_{x2}$
and $\epsilon_{z2}$ usually plays a pivotal role, 
for extremely thin HMM widths this is not the case. 
Indeed in the regime of small $\phi_2$,
Eq.~(\ref{d04}) simplifies to,
\begin{eqnarray}
r = \frac 
{ {\hat k}_z + i 2\pi\hat{\tau} (1-{\hat k}_x^2/\epsilon_{z2})}
{-{\hat k}_z + i 2\pi\hat{\tau} (1-{\hat k}_x^2/\epsilon_{z2})},
\end{eqnarray} 
which is independent of $\epsilon_{x2}$. Here we have introduced the dimensionless thickness:  $\hat{\tau}\equiv \omega\tau/c$.
Setting the numerator of $r$ 
to zero gives
the angle,  $\theta_c$,  where  the reflectance vanishes:
\begin{eqnarray}
\theta_c=\cos^{-1}\bigl[\bigl(i \epsilon_{z2}+
\sqrt{ (2\hat{\tau})^2(1-\epsilon_{z2})-\epsilon_{z2}^2}\,\,
\bigr)/({2\hat{\tau}})\bigr].
\label{thetac}
\end{eqnarray}
In Fig.~\ref{fig2}(a),  for the
incident wavelength of
$1.601$$\mu$m,
the approximate 
absorption angles
are found from taking the real part of
Eq.~(\ref{thetac}), 
giving, $\theta_c\approx$ $79^\circ$,$66^\circ$,$52^\circ$,$28^\circ$,
and $11^\circ$, in order of increasing $\tau$. 
Deviations in the angle predicted from Eq.~(\ref{thetac})
arise for larger $\tau$ as higher order corrections  are needed.
As the thickness $\tau$ decreases, 
near-perfect absorption shifts towards grazing incidences,
in agreement with Eq.~(\ref{thetac}) where as
$\tau\rightarrow0$, $\theta_c \rightarrow \pi/2$.
For the type-2 HMM, similar trends
are seen in Fig.~\ref{fig2}(b),  where 
$\lambda=1.59\mu$m and the near-perfect
absorption angles were  
found to agree well with Eq.~(\ref{thetac}).
It is apparent that for a type-2 HMM, the angular range of near-perfect absorption
exhibits a greater sensitivity to $\tau$ than the type-1 case shown.
In both cases (a) and (b), near-perfect absorption 
can be controlled over nearly the whole angular range by
properly choosing the effective material thicknesses.
When calculating  the 
regions of high absorption,
$\Im (\epsilon_{z2})$ plays a significant role when
$\Re (\epsilon_{z2})\approx 0$. This is consistent with
anisotropic leaky-wave structures \cite{klaus2}
and coherent perfect absorbers \cite{klaus3}.
Although more difficult to achieve in practice, 
subwavelength isotropic slabs where the permittivity and permeability simultaneously vanish,
can exhibit perfect absorption for small loss and a perfectly conducting metal backing \cite{he}. 
Additional control of  absorption  
may also be possible with the introduction of 
gain media  \cite{ni}.

\begin{figure}
\centerline{\includegraphics[width=8.0cm]{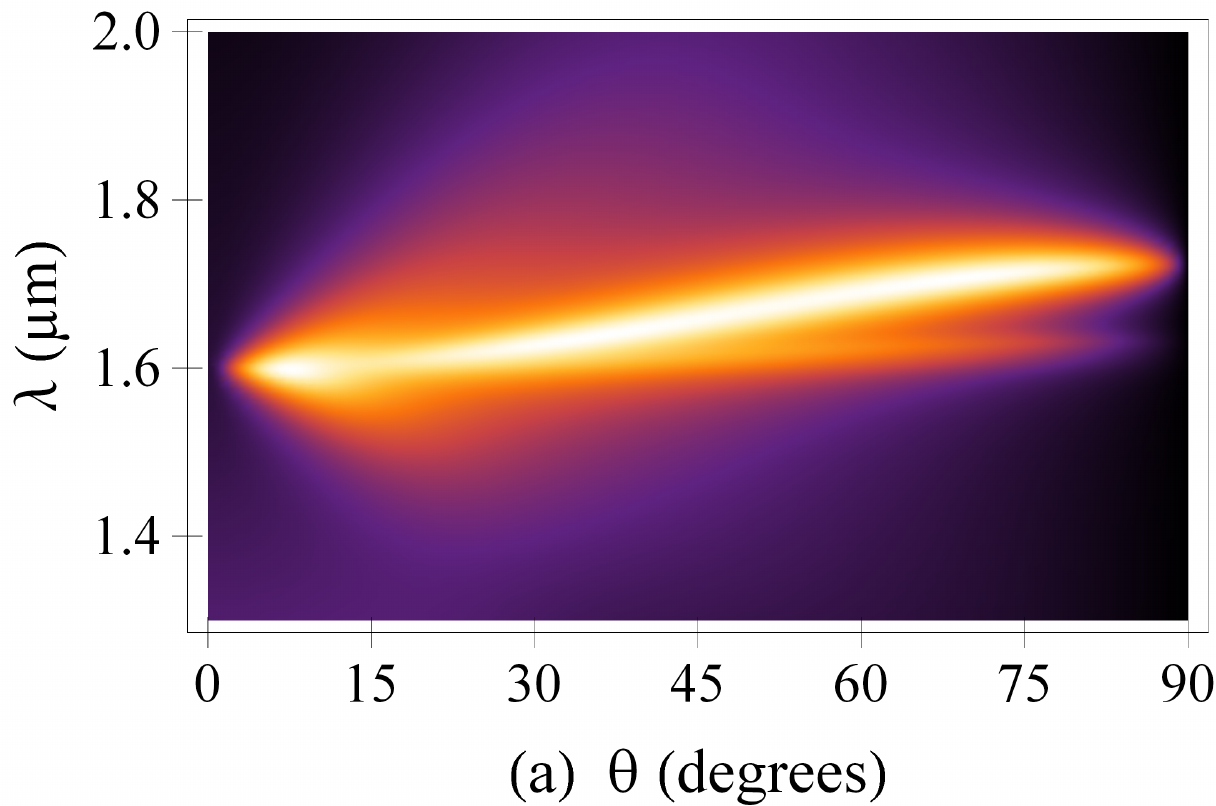} }
\centerline{\includegraphics[width=8.0cm]{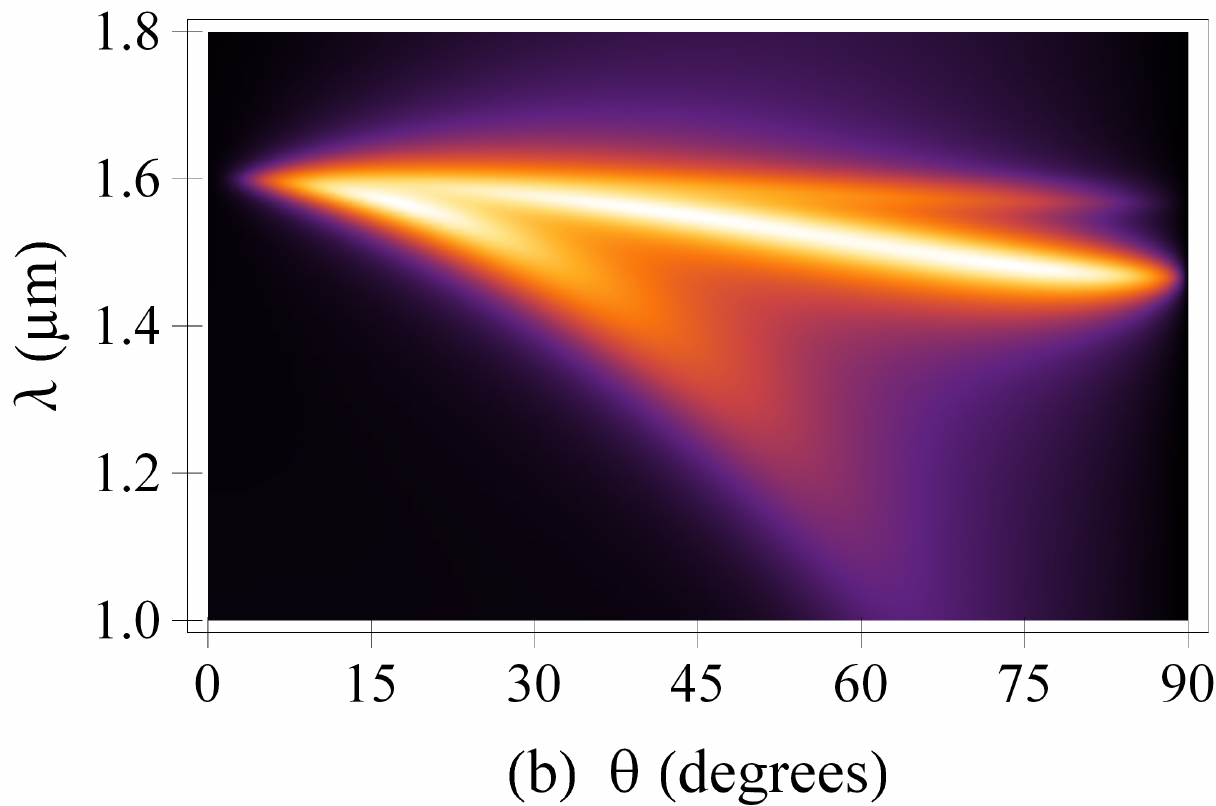} } 
\caption{Density plots showing absorption as a function of incident wavelength
$\lambda$ and angle $\theta$. Bright regions correspond to high absorption.
The HMM thickness in both plots is $\tau=0.16 \mu$m. The characteristic wavelength, 
$\lambda_{z} = 1.6 \mu$m separates the HMM regions according to
(a) type-1: $\epsilon_{x2} >0$ and $\epsilon_{z2} <0$ for $\lambda>1.6 \mu$m, and (b)
type 2: $\epsilon_{x2} <0$, and $\epsilon_{z2} >0$ for $\lambda<1.6 \mu$m.
Thus we find that when the metamaterial is effectively hyperbolic,
absorption can be strongly enhanced.
}  
 \label{fig3}
\end{figure}

\begin{figure} 
\centerline{\includegraphics[width=7cm]{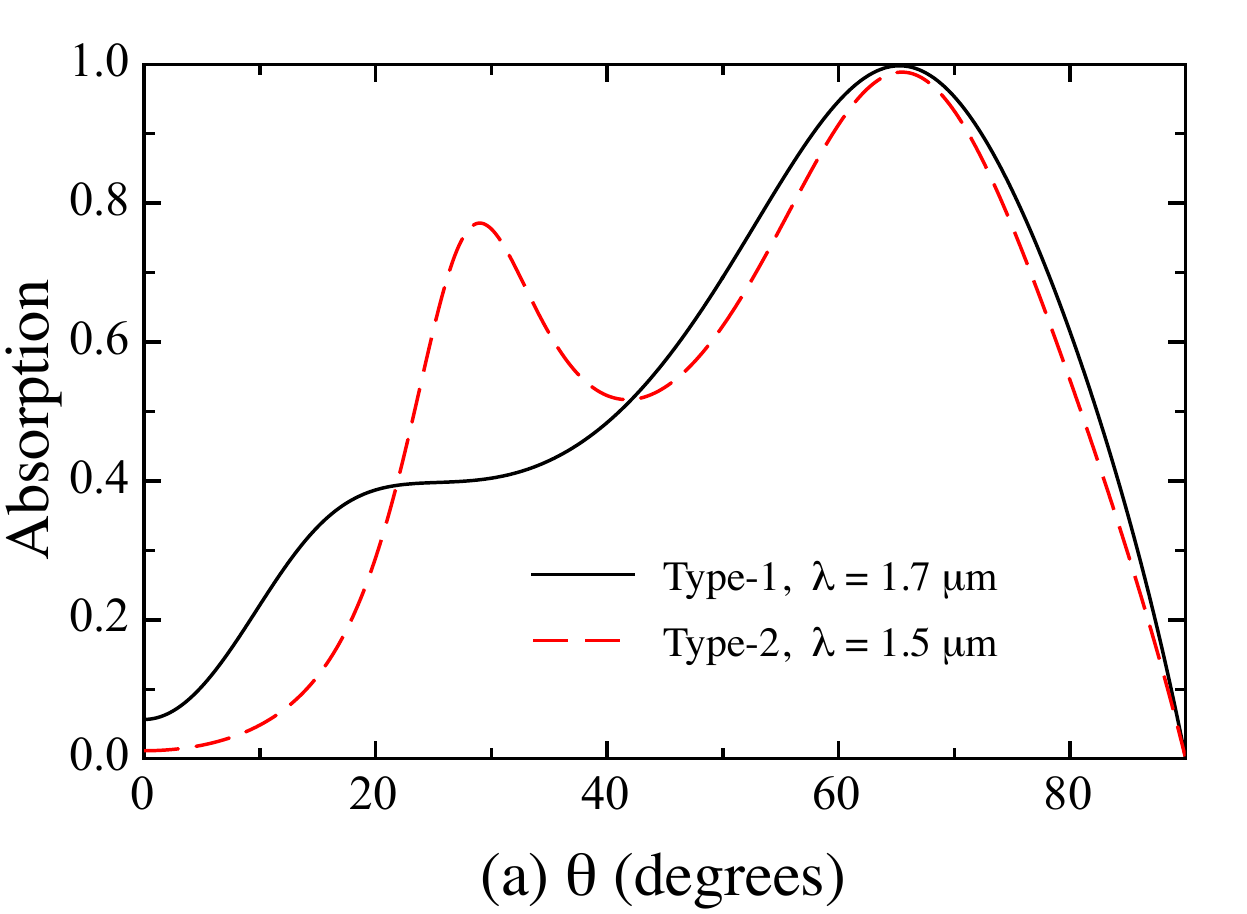} }
\centerline{\includegraphics[width=7cm]{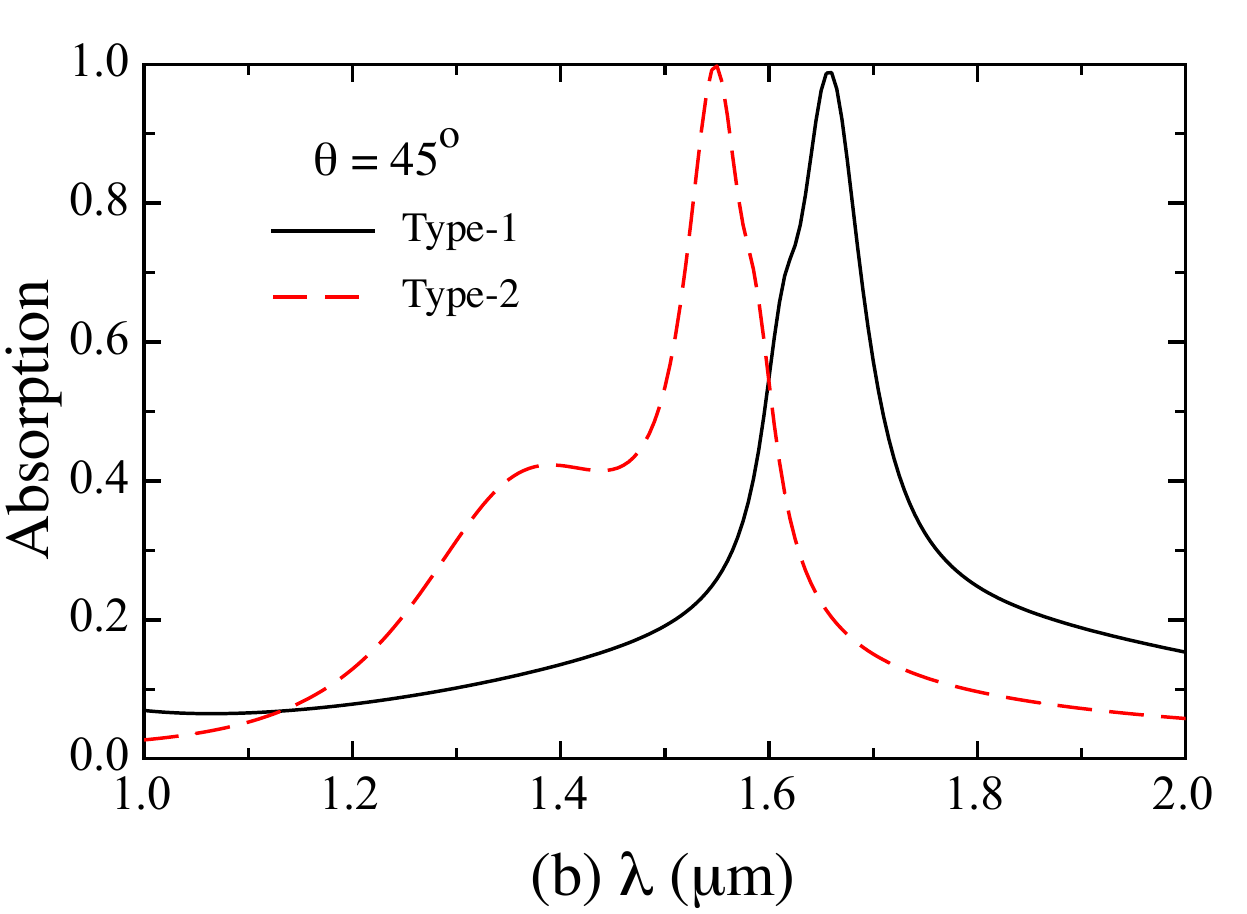} } 
\caption{Absorption
as a function of incident angle $\theta$ (a) and
wavelength $\lambda$ (b) extracted from the
high absorption regions of the
density plots in Fig.~\ref{fig3}(a) and (b).
}  
 \label{fig4}
\end{figure}
Next we investigate how varying the
wavelength of the source beam affects the absorption features.
In Figs.~\ref{fig2}(c) and (d) the thickness $\tau$ 
is set to $0.16\mu$m for both the type-1 and type-2 HMM cases respectively. 
For the type-1 HMM (panel c), 
as $\lambda$ increases 
beyond $\lambda_z$, 
the wavelength-dependent $\epsilon_{z2}$ shifts so that its
real part becomes more negative.
The corresponding absorption peaks then migrate towards $\theta=90^\circ$.
The opposite trend occurs for the type-2 case, where increasing $\lambda$
from $\lambda=1.5\mu$m
causes $\Re(\epsilon_{z2})$ (which is positive at this wavelength) to approach zero.
Consequently, the observed double-peaked absorption shifts towards normal incidence,
consistent with the trends above and Eq.~(\ref{thetac}),  where
as  $\lambda \rightarrow \lambda_z$ (and hence 
$\Re(\epsilon_{2z})\rightarrow 0$), the angle of 
near-perfect absorption tends to zero.
It is worth noting that if the HMM is replaced 
by an isotropic metallic layer like silver, 
the condition where the
permittivity is near zero 
is consistent with the generation
of bulk longitudinal collective 
oscillations of the free electrons. 
This type of excitation can produce moderate (but less than $100\%$)
absorption when there is minimal intrinsic
material loss.

For the case of
vacuum superstrate and substrate, 
Eq.~(\ref{e15}) reveals 
that when $\sin\phi_2 = 0$, then 
$R=0$.
If on the other hand, both substrate and superstrate are perfectly conducting, 
then setting the denominator of (\ref{e15}) equal to zero also yields $\sin\phi_2 = 0$,
which coincides with the dispersion relation for guided waves in an HMM layer.   
Equation (\ref{d03.2}) shows that
when $n=0$, 
${\hat k}_x^2 = \epsilon_{z2}\mu_{y2}$ and ${\hat k}_z^2 = 0$,
corresponding to a 
TEM mode which is essentially a plane wave 
confined to propagate in the $x$-direction. Thus if $\phi_2 = {\hat k}_{z2}\hat{\tau} = n\pi$, 
this
assertion is valid if $\phi_2 << n\pi$. If however $\epsilon_{z2}/\epsilon_{x2} < 0$, 
Eq.~(\ref{d03.2}) reveals that there is no guided mode cutoff  for ${\hat k}_x^2$.

To present a global view of
the parameter space in which our anisotropic structure can absorb
unusually 
large portions of incident energy, we present in Figs.~\ref{fig3}(a) and (b),
2-D density plots that map the absorption versus  $\lambda$ and  $\theta$.
The HMM thickness is fixed at $\tau$$=$ $0.16\mu$m, 
as in Figs.~\ref{fig2}(c) and (d).
In Fig.~\ref{fig3}(a) $\epsilon_{x2} = (4, 0.1)$,
so that the HMM region
where $\Re( \epsilon_{z2})<0$ 
corresponds to
$\lambda$ $>$ $\lambda_z $
(recall that $\lambda_z$=$1.6\mu$m).
Similarly for (b), $\epsilon_{x2} = (-4, \ 0.1)$, 
and thus 
the HMM region there corresponds to 
$\lambda$ $<$ $1.6 \mu$m.
Figs.~\ref{fig4}(a) and (b)
are slices from Figs.~\ref{fig3}(a) and (b).  
In Fig.~\ref{fig4}(a) near-perfect absorption occurs at $\theta = 65^\circ$ for both HMM types.
For  $\lambda = 1.7\mu$m,
$\Re(\epsilon_{x2}) = 4$ and $\Re(\epsilon_{z2}) = -0.128$ 
corresponding to a Type-1 HMM. 
For $\lambda = 1.5\mu$m, $\Re(\epsilon_{x2}) = -4$ and $\Re(\epsilon_{z2}) = 0.121$, 
corresponding to a Type-2 HMM. 
In Fig.~\ref{fig4}(b), the 
Type-1 absorption peak occurs 
at $\lambda = 1.66\mu$m, where $\Re(\epsilon_{z2}) = 
-0.076$,
and the Type-2 case
peaks at $\lambda = 1.55\mu$m, where $\Re(\epsilon_{z2}) = 0.062$. 

Further insight into this anomalous absorption can
be gained from studying the 
balance of energy \cite{jackson}. For our structure 
and material parameters, it suffices to compute,
\begin{eqnarray}
\label{edotjt}
\frac{4 \pi}{c}\int_V  dv \ {\bm E}\cdot{\bm J}^* = -\int_V dv \ \nabla\cdot\left({\bm E}\times{\bm H}^*\right)  
 - \frac{i\omega}{c}\int_V dv \left[\epsilon^*_{x2}|E_{x2}|^2+\epsilon^*_{z2}|E_{z2}|^2-|H_{y2}|^2 \right].
\end{eqnarray}
Since we have incorporated the conductive part of the HMM into the dielectric response, the $\bm J$ term is absent. 
In all of the near-perfect absorption examples investigated here, 
evaluation of Eq.~(\ref{edotjt}) confirmed
that the net energy flow into 
the HMM volume, $V$, is converted into heat.

\begin{figure} 
\centerline{\includegraphics[width=7.4cm]{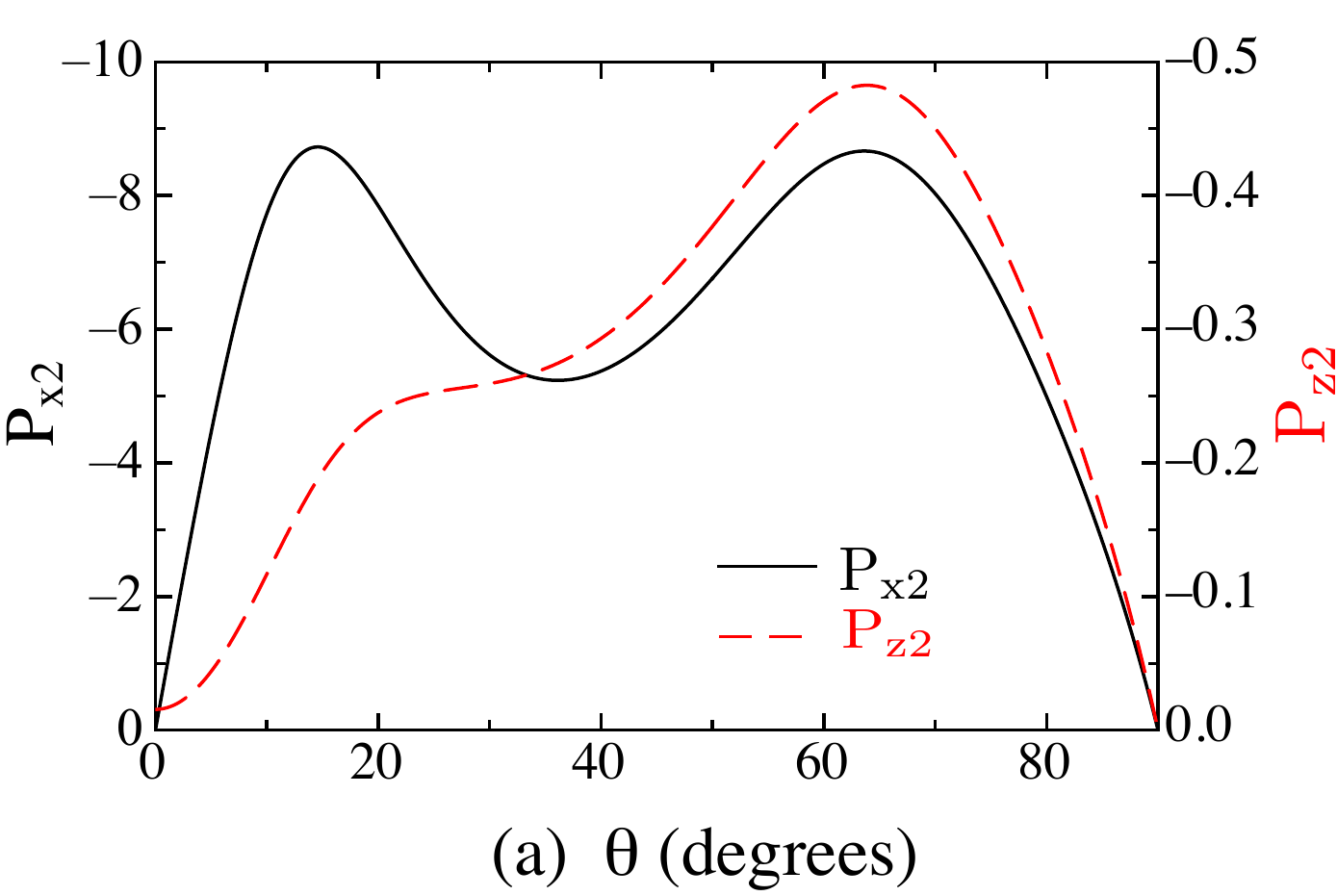}}
\centerline{\includegraphics[width=7.4cm]{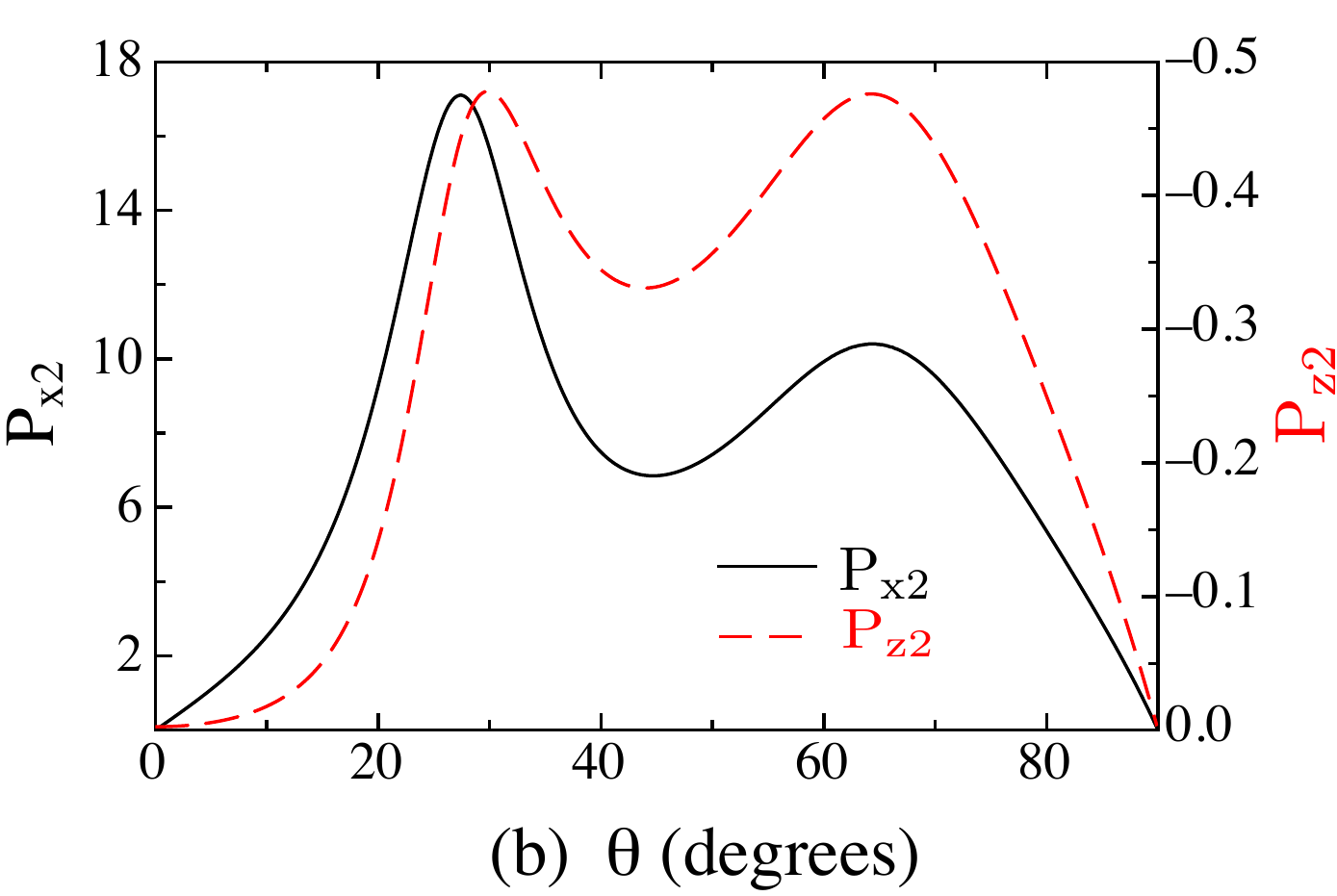}}
\caption{ 
The power ${\bm P}$ in the HMM, 
normalized by the incident power in the 
$z$-direction, and plotted as a function of $\theta$.
In (a) $\lambda = 1.7$ $\mu$m (type-1 HMM) and
in (b) $\lambda = 1.5$ $\mu$m (type-2 HMM).  
The material parameters are the same as in Fig.~\ref{fig4}(c). 
Panel (a) reveals that 
energy flow parallel to the interface ($P_{x2}$) in the type-1
HMM is negative, 
which is opposite
that of the vacuum region containing the incident beam.
}  
\label{fig5}
\end{figure}
To explore further the behavior of the energy flow,
we 
present in Fig.~\ref{fig5}
the average power ${\bm P}$ in the HMM
as a function
of $\theta$
for the two cases in Fig.~\ref{fig4}(a).
Thus, panel (a) is for $\lambda = 1.7$$\mu$m (type-1 HMM), 
and  panel (b) corresponds
to $\lambda = 1.5$$\mu$m (type-2 HMM).
The average power along 
the $x$ and $z$ directions,  $P_{x2}$ and $P_{z2}$,
is found from 
averaging the corresponding components of the 
Poynting vector
over the HMM region (see Eqs.~(\ref{s2})).
It is  evident that
the direction of energy flow depends on
the sign of $\epsilon_{z2}$ (or equivalently whether $\lambda$ is
above or below $\lambda_z$).
The component of $\bm P$ normal to the interfaces ($P_{z2}$) must always
have the same sign on both sides of the interface \cite{smith}.
Its direction parallel to the interface ($P_{x2}$) however
can be  negative if the HMM is of type-1, as seen in
Fig.~\ref{fig5}(a),
and is clearly opposite in direction to 
$k_x$,
which is always positive.
This manifestation of ``negative refraction"
was discussed in the context of uniaxially anisotropic media \cite{lindell},
certain nanowire structures \cite{liu},
and observed in ZnO-based multilayers \cite{naik}.
Comparing the peaks in panels (a) and (b) with Fig.~\ref{fig4}(a), 
we see the correlation with the angles
where
near-perfect absorption occurs and those 
where  $|P_{z2}|$ is maximal.

\section{Conclusion}
In conclusion, we have investigated the absorption
properties of both type-1 and type-2 hyperbolic metamaterials.
We found that HMMs can absorb significantly higher amounts of
electromagnetic energy compared to their conventional 
counterparts, where $\Re(\epsilon_{x2})$ and $\Re(\epsilon_{z2})$
are both of the same sign. 
Our results show that the incident beam can couple to
the HMM structure without recourse for
a second compensating layer.
We also revealed
that the condition $\Re(\epsilon_z) \approx 0$
leads to near-perfect absorption 
over a range of frequencies, angles of incidence, and subwavelength structure thicknesses,
making the proposed structures experimentally achievable.
Alternate methods exist to achieve perfect absorption, including
periodic layers of silver and conventional dielectrics  that 
depending on the direction of incident wave propagation and loss, 
can exhibit anisotropic behavior 
that  cancels the reflected and transmitted waves simultaneously  \cite{yang}. 
Our HMM with 
metallic backing is a different configuration in which 
no energy can be transmitted,
and the inherently finite width of the structure
means that there are no
Bloch wave excitations.
Arrays of metal-dielectric films
can serve as an effective HMM waveguide taper,
resulting in light localization and enhanced absorption \cite{rainbow}, 
however, the modes responsible for ``slow-light" are 
very sensitive to the presence of loss \cite{lu}.

When the incident wavelength results
in the dielectric response of the metamaterial
possessing
a nearly vanishing component 
of
the permittivity, 
contributions from nonlinear effects and/or  
spatial dispersions can become important. 
Nonlinear effects can in this case generate 
interesting
phenomena such as two-peaked or flat solitons \cite{rizza3}, 
as well as
additional venues for second- and third-harmonic generation \cite{vince},
and guided waves
whose
Poynting vector undergoes localized reversal  \cite{rizza4}. 
Since the nonlinear part of the dielectric response can now be of the
same order as the (small) linear part, the
transmissivity can exhibit directional hysteresis behavior \cite{rizza5}.
Spatial dispersion can moreover lead to the appearance of additional EM waves, as was
reported for nanorods \cite{pol}.
For metal-dielectric structures, nonlocality
arising from the excitation of surface plasmons can
also lead to significant corrections \cite{orlov} to conventional
effective medium theories \cite{berg}.

\appendix
\section*{Appendix: Poynting's Theorem}  \label{app}
In this section
we present the details on how the EM fields are
straightforwardly calculated 
for determining the reflectance and 
energy flow in HMM structures.
We have considered in this paper
diagonally anisotropic
HMM layers ($\epsilon_x, \epsilon_z, \mu_y$).  
We also assume that EM
wave propagation and polarization is 
in the $x$-$z$ plane.  
The wave equation for $E_x$ is thus,
\begin{eqnarray}
\label{e01}
\frac{\partial^2 E_x }{\partial z^2} + \left[\left(\frac{\omega}{c}\right)^2 \epsilon_{x}\mu_{y} - \left(\frac{\epsilon_{x}}{\epsilon_{z}}\right) k_x^2 \right]  E_x = 0.
\end{eqnarray}
Taking into account that $\nabla \cdot {\bm D} = \nabla \cdot {\bm B} = 0$, 
this yields the electric field solutions in their respective media as,
\begin{eqnarray}
\label{e02a}
{\bm E}_1 &=& \left[ A\left\{{\hat x} + {\hat z}\left(\frac{{\hat k}_x \epsilon_{x1}}{{\hat k}_{z1} \epsilon_{z1}}\right) \right\} e^{-i k_{z1} z} \right] e^{i k_x x}, \\
\label{e02b}
{\bm E}_2 &=&  \left[ 
G \left\{
{\hat x} - {\hat z}\left(\frac{{\hat k}_x \epsilon_{x2}}{{\hat k}_{z2} \epsilon_{z2}}\right) 
\right\}
e^{i k_{z2} z }
+ F \left\{
{\hat x} + {\hat z}\left(\frac{{\hat k}_x \epsilon_{x2}}{{\hat k}_{z2} \epsilon_{z2}}\right)
\right\}
e^{-i k_{z2} z} 
\right]
e^{i k_x x}, \\
\label{e02c}
{\bm E}_3 &=&  \left[ 
C \left\{
{\hat x} - {\hat z}\left(\frac{{\hat k}_x \epsilon_{x3}}{{\hat k}_{z3} \epsilon_{z3}}\right) 
\right\}
e^{i k_{z3} z}
+ I \left\{
{\hat x} + {\hat z}\left(\frac{{\hat k}_x \epsilon_{x3}}{{\hat k}_{z3} \epsilon_{z3}}\right)
\right\}
e^{-i k_{z3} z } 
\right]
e^{i k_x x}.
\end{eqnarray}
Similarly, the components of the magnetic field are written,
\begin{eqnarray}
\label{e02d}
{\bm H}_1 &=& \ -{\hat y}\left(\frac{\epsilon_{x1}}{{\hat k}_{z1}}\right) A  e^{-i k_{z1} z} e^{i k_x x}, \\
\label{e02e}
{\bm H}_2 &=& \ {\hat y}\left(\frac{\epsilon_{x2}}{{\hat k}_{z2}}\right)\Big[G e^{i k_{z2} z } - F e^{-i k_{z2} z } \Big]e^{i k_x x}, \\
\label{e02fc}
{\bm H}_3 &=& \ {\hat y}\left(\frac{\epsilon_{x3}}{{\hat k}_{z3}}\right)\Big[C e^{i k_{z3} z} - I e^{-i k_{z3} z} \Big]e^{i k_x x}.
\end{eqnarray}
The $I$ terms represent the incident field. 
The quantities ${\hat k}_{zj}$ 
and $\phi_j$
are defined in Eqs.~(\ref{phij}) and (\ref{kzj}).
Utilizing
matching boundary conditions for the tangential components of the 
electric and magnetic fields permits calculation of the coefficients,
\begin{eqnarray}
A&=&\frac
{4 e^{-i\phi_3}{\hat k}_{z1}{\hat k}_{z2} \epsilon_{x2}\epsilon_{x3}}
{{\cal G}^-{\cal F}^-e^{i\phi_2}+{\cal G}^+{\cal F}^+e^{-i\phi_2}}; \quad 
C=  \beta\left[\frac
{{\cal G}^-{\cal F}^+e^{i\phi_2}+{\cal G}^+{\cal F}^-e^{-i\phi_2}}
{{\cal G}^-{\cal F}^-e^{i\phi_2}+{\cal G}^+{\cal F}^+e^{-i\phi_2}}\right], \\
F&=& \frac
{2 e^{-i\phi_3}{\hat k}_{z2}\epsilon_{x3} {\cal G}^+}
{{\cal G}^-{\cal F}^-e^{i\phi_2}+{\cal G}^+{\cal F}^+e^{-i\phi_2}}; \quad
G=\frac
{-2 e^{-i\phi_3}{\hat k}_{z2}\epsilon_{x3} {\cal G}^-}
{{\cal G}^-{\cal F}^-e^{i\phi_2}+{\cal G}^+{\cal F}^+e^{-i\phi_2}},
\end{eqnarray}
were $\beta$ is given in Eq.~(\ref{beta}). 
We also define,
\begin{eqnarray}
{\cal F}^\pm &=& {\hat k}_{z3}\epsilon_{x2} \pm {\hat k}_{z2}\epsilon_{x3}, \\
{\cal G}^\pm &=& {\hat k}_{z2}\epsilon_{x1} \pm {\hat k}_{z1}\epsilon_{x2},
\end{eqnarray}
where $\epsilon_{xj}$  describe the media for regions $j = 1,2,3$,
and $k_{zj}$ is defined in Eq.~(\ref{kzj}).
The caret symbol signifies that 
wavenumber components $k_x$ and $k_{zj}$ have been normalized to $\omega/c$.  
In general, ${\hat k}_x$ can be any value, 
but for the case of an incident plane wave in vacuum, ${\hat k}_x = \sin\theta$. 

For time-harmonic fields, 
consider now the integral,
\begin{eqnarray}
\label{edotj}
\frac{1}{2}\frac{4 \pi}{c}\int_V dv \ {\bm E}\cdot{\bm J}^* = -\frac{1}{2}\int_V dv \ \nabla\cdot\left({\bm E}\times{\bm H}^*\right)   
- \frac{i\omega}{2c}\int_V dv \left[{\bm E}\cdot{\bm D}^* - {\bm H}^*\cdot{\bm B} \right],
\end{eqnarray}
where we have used,
\begin{eqnarray}
\label{misc}
\nabla \times {\bm H} = \frac{4\pi}{c}{\bm J}-\frac{i\omega}{c}{\bm D} \quad {\rm ;} \quad \nabla \times {\bm E} = \frac{i\omega}{c}{\bm B}.
\end{eqnarray}
The media are diagonally anisotropic with ${\bm D} = {{\bm \epsilon}}\cdot{\bm E}$ and ${\bm B} = {\bm{\mu}}\cdot{\bm H}$. 
Inserting Eqs.~(\ref{e02b}) and (\ref{e02e}) into Eq.~(\ref{edotj}) yields the following energy conservation relationships,
\begin{eqnarray}
\label{exdx}
\frac{i\omega}{2c}\int_0^{\tau} \ dz \ {E_{2x}}{D^*_{2x}} & = &\ \frac{i\epsilon^*_{2x}}{2} \Bigl[|G|^2 
\Bigl( \frac{ {\rm e}^{-2\tau\Im({k_{2z}})} - 1 } { {-2\Im{({\hat k}_{2z})}} } \Bigr) 
+ |F|^2  \Bigl( \frac{ {\rm e}^{2\tau\Im({k_{2z}})} - 1}{{2\Im{({\hat k}_{2z})}} }\Bigr) \nonumber \\
&+& 2\Re\Bigl\{ GF^* \Bigl( \frac{ {\rm e}^{2i\tau\Re({k_{2z}})} - 1 } { {2i\Re{({\hat k}_{2z})}} } \Bigr)\Bigr\} \Bigr],
\end{eqnarray}
\begin{eqnarray}
\label{ezdz} 
\frac{i\omega}{2c}\int_0^{\tau} \ dz \ {E_{2z}}{D^*_{2z}} & = &
\ \frac{i\epsilon^*_{2z}}{2} \bigg| \frac{{\hat k}_x \epsilon_{2x}}{{\hat k}_{2z} \epsilon_{2z}} \bigg|^2 \Bigl[|G|^2 
\Bigl( \frac{ {\rm e}^{-2\tau\Im{(k_{2z})}} - 1 } { {-2\Im{({\hat k}_{2z})}} } \Bigr) \nonumber \\ 
&+& |F|^2  \Bigl( \frac{ {\rm e}^{2\tau\Im{(k_{2z})}} - 1}{{2\Im{({\hat k}_{2z})}} }\Bigr) 
- 2\Re\Bigl\{ GF^* \Bigl( \frac{ {\rm e}^{2i\tau\Re({k_{2z}})} - 1 } { {2i\Re{({\hat k}_{2z})}} } \Bigr)\Bigr\} \Bigr],
\end{eqnarray}
\begin{eqnarray}
\label{hyby}
\frac{i\omega}{2c}\int_0^{\tau} \ dz \ {H^*_{2y}}{B_{2y}} && = \ \frac{i\mu_{2y}}{2} \bigg| \frac{\epsilon_{2x}}{{\hat k}_{2z}} \bigg|^2 \Bigl[|G|^2 
\Bigl( \frac{ {\rm e}^{-2\tau\Im{(k_{2z})}} - 1 } { {-2\Im{({\hat k}_{2z})}} } \Bigr) \nonumber \\ 
&&+ |F|^2  \Bigl( \frac{ {\rm e}^{2\tau\Im{(k_{2z})}} - 1}{{2\Im{({\hat k}_{2z})}} }\Bigr) 
- 2\Re\Bigl\{ GF^* \Bigl( \frac{ {\rm e}^{2i\tau\Re{(k_{2z})}} - 1 } { {2i\Re{({\hat k}_{2z})}} } \Bigr)\Bigr\} \Bigr],
\end{eqnarray}
where the $x$ and $y$ integrations over $V$ are omitted.

Finally, the time-averaged Poynting vector in $V$ is ${\bm S} = {\bm E} \times {\bm H}^*/2$, giving the result,
\begin{eqnarray}
\label{s2}
S_{2x}(z) &=& \ \Bigl(\frac{{\hat k}_x}{\epsilon_{z2}}\Bigr)\bigg|\frac{\epsilon_{x2}}{{\hat k}_{z2}} \bigg|^2 \bigg[|G|^2 {\rm e}^{-2z\Im (k_{z2})} + |F|^2 {\rm e}^{2z\Im (k_{z2})} 
-2 \Re \Bigl( G F^* {\rm e}^{2iz\Re (k_{z2})}\Bigr) \bigg], \\
S_{2z}(z) &=& \ \Bigl(\frac{\epsilon_{x2}}{{\hat k}_{z2}} \Bigr)^* \bigg[|G|^2 {\rm e}^{-2z\Im (k_{z2})} - |F|^2 {\rm e}^{2z\Im (k_{z2})}  
-2i \Im \Bigl( G F^* {\rm e}^{2iz\Re (k_{z2})}\Bigr) \bigg].
\end{eqnarray}

\end{document}